\documentstyle[12pt,epsfig,amsmath]{article}
\def\bge{\begin{equation}}
\def\ene{\end{equation}}
\def\bg{\begin{eqnarray}}
\def\en{\end{eqnarray}}

\begin{document}
\title{The role of $\Delta^{++*}(1620)$ resonances in
 $pp \to nK^+\Sigma^+$ reaction and its important implications}
\author{
Ju-Jun Xie~$^{1,2)}$\thanks{xiejujun@mail.ihep.ac.cn},
Bing-Song Zou$^{1,2,3)}$\thanks{zoubs@mail.ihep.ac.cn}  \\
{1) Institute of High Energy Physics, CAS, P.O.Box 918 (4), Beijing 100049 } \\
2) Graduate University, Chinese Academy of Sciences, Beijing
100049 \\
3) Center of Theoretical Nuclear Physics, National Laboratory of\\
Heavy Ion Accelerator, Lanzhou 730000 }
\date{January 9, 2007}
\maketitle

\begin{abstract}
The $pp \to nK^+\Sigma^+$ reaction is a very good isospin $3/2$
filter for studying $\Delta^{++*}$ decaying to $K^+\Sigma^+$. With
an effective Lagrangian approach, contributions from a previous
ignored sub-$K^+\Sigma^+$-threshold resonance
$\Delta^{++*}(1620)1/2^-$ are fully included in addition to those
already considered in previous calculations. It is found that the
$\Delta^{++*}(1620)1/2^-$ resonance gives an overwhelmingly
dominant contribution for energies very close to threshold, with a
very important contribution from the t-channel $\rho$ exchange.
This solves the problem that all previous calculations seriously
underestimate the near-threshold cross section by order(s) of
magnitude. Many important implications of the results are
discussed.
\end{abstract}

\section{Introduction}

The spectrum of isospin 3/2 $\Delta^{++*}$ resonances is of
special interest since it is the most experimentally accessible
system composed of 3 identical valence quarks. However, our
knowledge on these resonances mainly comes from old $\pi N$
experiments and is still very poor~\cite{pdg2006,manley,capstick}.
Many model-predicted $\Delta^*$ resonances have not been
established or ever observed~\cite{capstick,riska}. A possible
reason is that these ``missing" baryon resonances have too weak
coupling to $\pi N$ to be observed in the $\pi N$ experiments.
Searching for these ``missing" $\Delta^*$ resonances from other
production processes is necessary. A possible new excellent source
for studying $\Delta^{++*}$ resonances is $pp \to nK^+\Sigma^+$
reaction, which has a special advantage for absence of
complication caused by $N^*$ contribution because of the isospin
and charge conversation.

At present, little is known about the $pp  \to nK^+\Sigma^+$
reaction. Experimentally there are only a few data points about
its total cross section versus energy~\cite{data,06cosy11}.
Theoretically a resonance model with an effective intermediate
$\Delta^{++*}(1920)$ resonance~\cite{tsushima,sibi,shyam06} and
the J\"{u}lich  meson exchange model~\cite{gas} reproduce the old
data at higher beam energy~\cite{data} quite well, but their
predictions for the cross sections close to threshold fail by
order of magnitude compared with very recent COSY-11
measurement~\cite{06cosy11}. Here we restudy this reaction by
examining various possible sources for the very strong
near-threshold enhancement.

In the resonance model with an effective intermediate
$\Delta^{++*}(1920)$ resonance~\cite{tsushima,sibi,shyam06}, the
$3/2^+$ $\Delta^{++*}(1920)$ resonance decays to $K^+\Sigma^+$ in
relative P-wave and is suppressed at lower energies. To reproduce
the near-threshold enhancement for the $pp  \to nK^+\Sigma^+$
reaction, a natural source could be some $1/2^-$ $\Delta^{++*}$
resonance(s) at lower energy decaying to $K^+\Sigma^+$ in relative
S-wave. In the J\"{u}lich meson exchange model~\cite{gas}, only
$\pi$ and K exchanges are included with $\pi^+p\to K^+\Sigma^+$
and $KN \to KN$ amplitudes taken from the relevant existing data.
Its failure at lower energy indicates that other meson exchange
might be important. Following the logic, we find a natural source
for the near-threshold enhancement of the $pp \to nK^+\Sigma^+$
reaction coming from $\rho^+p\to \Delta^{++*}(1620) (1/2^-)\to
K^+\Sigma^+$. The well established $\Delta^{++*}(1620)$ has an
unusual large decay branching ratio to the $p\rho^+$
\cite{pdg2006,dytman} and its coupling to $K^+\Sigma^+$ can be
obtained from its coupling to $\pi^+p$ with SU(3) symmetry.
Possible $n$-$\Sigma^+$ final state interaction are also studied.

In next section, we will give the formalism and ingredients for
our calculation. Then numerical results and discussion are given
in Sect.3.

\section{Formalism and ingredients}

\begin{figure}[htpb]
\centerline{\epsfig{file=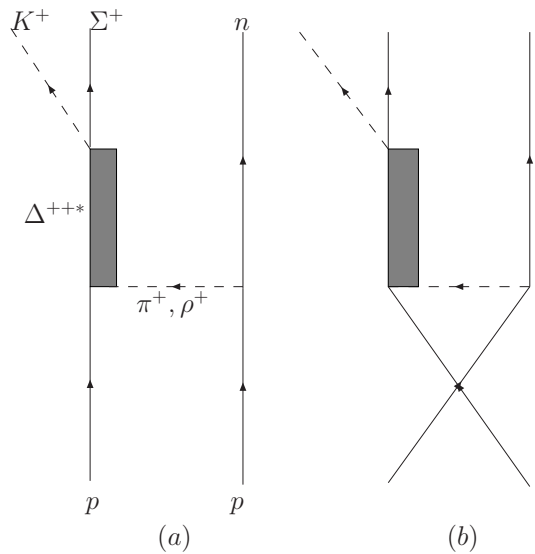,width=7cm}} \vspace{0cm}
\caption{Feynman diagrams for $pp \to n K^+\Sigma^+$ reaction. }
\label{diagram}
\end{figure}

The basic Feynman diagrams for the $pp \to n K^+ \Sigma^+$
reaction are depicted in Fig.~\ref{diagram}. Besides the
$\Delta^{++*}(1920)3/2^+$ considered in the previous resonance
model \cite{tsushima,sibi,shyam06}, $\Delta^{++*}(1620)1/2^-$
resonance is added in our calculation. Both $\pi^+$ and $\rho^+$
exchanges are considered for the production of $\Delta^{++*}$
resonances. A Lorentz covariant orbital-spin(L-S) scheme
\cite{zouprc03} are used for the effective $\Delta^{*}\pi N$,
$\Delta^{*}\rho N$, and $\Delta^{*} K \Sigma$ vertices.

\subsection{Meson-Baryon-Baryon (Resonances) vertices}

The relevant effective Lagrangian densities for the meson-NN
vertices are the standard ones \cite{tsushima}:
\begin{equation}
{\cal L}_{\pi N N}  = -i g_{\pi N N} \bar{N} \gamma_5 \vec\tau
\cdot \vec\pi N, \label{sp}
\end{equation}
\begin{equation}
{\cal L}_{\rho N N} = -g_{\rho N N}
\bar{N}(\gamma_{\mu}+\frac{\kappa}{2m_N} \sigma_{\mu \nu}
\partial^{\nu})\vec\tau \cdot \vec\rho^{\mu} N. \label{sr}
\end{equation}
And the relevant off-shell form factors for the $\pi^+$-NN and
$\rho^+$-NN vertices are taken as the same as the well known Bonn
potential model~\cite{mach}:
\begin{equation}
F^{NN}_M(q^2_M)=(\frac{\Lambda^2_M-m_M^2}{\Lambda^2_M- q_M^2})^n
\end{equation}
with n=1 for $\pi^+$-meson and n=2 for $\rho^+$-meson. $q_M$,
$m_M$ and $\Lambda_M$ are the 4-momentum, mass and cut-off
parameter of the exchanged-meson ($M$), respectively. Parameters
are taken as commonly used ones \cite{mach,tsushima,sibi}: $g_{\pi
NN}^2/4\pi = 14.4$, $\Lambda_{\pi}$ = 1.3 GeV, $g_{\rho
NN}^2/4\pi$ = 0.9,  $\Lambda_{\rho}$ = 1.85 GeV, and $\kappa =
6.1$.

To calculate the amplitudes for diagrams in Fig.~\ref{diagram}
with resonance model, we also need to know interaction vertices
involving $\Delta^*$ resonances. In Ref.~\cite{zouprc03}, a
Lorentz covariant orbital-spin scheme for $N^* NM$ couplings has
been described in detail. The scheme can be easily extended to
describe the $\Delta^* \pi N$, $\Delta^* \rho N$ and $\Delta^* K
\Sigma$ couplings that appears in the Feynman diagrams showing in
Fig.~\ref{diagram}.

Firstly, for $\Delta^{++*} \to \pi^+$p, it is well known that
there is only one possible L-S coupling for the $\pi^+$p final
state of each $\Delta^{++*}$ decay. Since the nucleon has
spin-parity $\frac{1}{2}^+$ and pion has spin-parity $0^-$, with
the parity and angular momentum conversation,
$\Delta^{++*}(1620)(\frac{1}{2}^-)$ and
$\Delta^{++*}(1920)(\frac{3}{2}^+)$ can only decay to $\pi^+$p in
S-wave and in P-wave, respectively.  The effective
$\Delta^{++*}\pi^+$p couplings in the covariant orbital-spin
scheme can be easily obtained as :
\begin{eqnarray}
\Delta^{++*}(1620)(\frac{1}{2}^-) \to \pi^+p &:& \hspace{1.0cm}
g_{\Delta^*(1620) N \pi}\overline{u}_{p}u_*, \label{1620g}\\
\Delta^{++*}(1920)(\frac{3}{2}^+) \to \pi^+p &:& \hspace{1.0cm}
g_{\Delta^*(1920) N \pi}\overline{u}_{p}u_{*\mu}p^\mu_\pi,
\label{1920g}
\end{eqnarray}
with $u_p$, $u_*$ and $u_{*\mu}$ as the Rarita-Schwinger spin wave
functions for the proton, $\Delta^{++*}(1620)$ and
$\Delta^{++*}(1920)$, respectively.

Secondly, for $\Delta^{++*} \to K^+\Sigma^+$, the effective
$\Delta^{++*} K^+ \Sigma^+$ couplings can be easily obtained with
SU(3) symmetry by simply replacing $\pi^+$ and $p$ with $K^+$ and
$\Sigma^+$, respectively.

Thirdly, for $\Delta^{++*} \to \rho^+ p$, unlike pion with spin 0,
here $\rho$ has spin 1. For a $\Delta^*$ with spin $\frac{1}{2}$
there are two independent L-S couplings conserving parity and
total angular momentum; for a $\Delta^*$ with spin larger than
$\frac{1}{2}$, there are three independent L-S couplings
\cite{zouprc03}. Here, in our calculation, we only consider the
$\Delta^{++*}(1620)(\frac{1}{2}^-)$ resonance which was found to
have large decay branch ratio to $\rho N$ in relative S-wave
\cite{pdg2006,dytman}. Following Ref.\cite{zouprc03}, the
effective coupling is obtained as
\begin{equation}
\Delta^{++*}(1620)(\frac{1}{2}^-) \to \rho^+ p:\hspace{0.5cm}
g_{\Delta^*(1900) N \rho}\overline{u}_{p} \gamma_5
(\gamma_{\mu}-\frac{p_{* \mu} \gamma^{\nu} p_{* \nu}}{p^2_*}) u_*
\varepsilon^{* \mu}, \label{1620rho}
\end{equation}
with $p_*$ and $\varepsilon^{\mu}$ the four momentum of the
$\Delta^*$ and the polarization vector of $\rho$ meson.

For the relevant vertices involving $\Delta^*$, since only S- and
P-waves are involved, we use the monopole form for the off-shell
form factors :
\begin{equation}
F_M(q)=\frac{\Lambda^{*2}_M-m_M^2}{\Lambda^{*2}_M- q_M^2},
\end{equation}
with $\Lambda^*_{\pi}$ = $\Lambda^*_{\rho}$ = 1.3 GeV.

\subsection{Coupling constants for $\Delta^{++*}$ resonances}

The relevant $\Delta^*$-baryon-meson coupling constants are
determined either from experimentally observed partial decay
widths or SU(3) symmetry.

The general formula for the partial decay width of a $\Delta^*$
resonance decaying to a nucleon and a pion is as the following
\begin{equation}
d \Gamma =\overline{|{\cal M}_{\Delta^*\to N\pi} |^2} (2\pi)^4
\delta^4(p_{\Delta^*}-p_N-p_\pi) \frac{d^3 p_N}{(2\pi)^3}
\frac{m_N}{E_N} \frac{d^3p_\pi}{(2\pi)^3} \frac{1}{2E_\pi}
\end{equation}
where ${\cal M}_{\Delta^*\to N\pi }$ represents the total
amplitude of the $\Delta^*$ decay to a nucleon and a pion, the
$p_{\Delta^*}$, $p_N$ and $p_\pi$ are the four momentum of the
three particles, $E_N$ and $E_\pi$ are the corresponding energies.

For $\Delta^{*}(1620) (\frac{1}{2}^-)$ $\to N \pi$,  the partial
decay width can be calculated with the amplitude given by
Eq.~(\ref{1620g}) to be
\begin{equation}
\Gamma_{\Delta^*(1620)\to N \pi}=\frac{g^2_{\Delta^*(1620) N
\pi}(m_N+E_N)p_N^{cm}}{4\pi M_{\Delta^*(1620)}} \label{1620d}
\end{equation}
with
\begin{equation}
p_N^{cm}=\sqrt{\frac{(M^2_{\Delta^*(1620)}-(m_N+m_{\pi})^2)
(M^2_{\Delta^*(1620)}-(m_N-m_{\pi})^2)}{4M^2_{\Delta^*(1620)}}},
\end{equation}
\begin{equation}
E_N=\sqrt{(p_N^{cm})^2+m^2_N} .
\end{equation}

For $\Delta^*(1920)(\frac{3}{2}^+) \to N \pi$, the partial decay
width can be calculated with the amplitude given by
Eq.(\ref{1920g}) to be
\begin{equation}
\Gamma_{\Delta^*(1920) N \pi}=\frac{g^2_{\Delta^*(1920) N
\pi}(m_N+E_N)(p_N^{cm})^3}{12 \pi M_{\Delta^*(1920)}}
\label{1920d}
\end{equation}
with $p_N^{cm}$ here the momentum of the nucleon in the rest frame
of $\Delta^*(1920)$.

For $\Delta^*\to K \Sigma$, the formulae are basically identical
as for $\Delta^* \to \pi N$ with the replacement of $\pi$ to kaon
and proton to $\Sigma$.

For $\Delta^{*}(1620) (\frac{1}{2}^-) \to N \rho \to N \pi \pi$,
the partial decay width of $\Delta^*(1620) \to N \rho$ in relative
S-wave via $\rho$ decay into two pions is given by
\begin{equation}
\Gamma_{\Delta^*(1620) N \rho} = \overline{|{\cal M}_{\Delta^*\to
N\rho\to N\pi\pi}|^2} \frac{d^3p_1}{(2 \pi)^3} \frac{m_1}{E_1}
\frac{d^3p_2}{(2 \pi)^3} \frac{1}{2 E_2} \frac{d^3p_3}{(2 \pi)^3}
\frac{1}{2 E_3} (2 \pi)^4 \delta^4(M_{\Delta^*(1620)}-p_1-p_2-p_3)
\end{equation}
where ${\cal M}_{\Delta^*\to N\rho\to N\pi\pi}$ represents the
total amplitude of $\Delta^{*}(1620) (\frac{1}{2}^-) \to N \rho
\to N \pi \pi$. $p_1$, $m_1$, and $E_1$ stand for the four
momentum, mass, and energy of the nucleon; $p_2$, $p_3$, and
$E_2$, $E_3$ stand for the four momentum and energy of the final
two pions, respectively. In the amplitude calculation, the
amplitude for $\rho \to \pi \pi$ is taken as ${\cal M}_{\rho \pi
\pi} = g_{\rho \pi \pi} e_{\mu}(\rho, m_{\rho})
(p^{\mu}_2-p^{\mu}_3)$ with $e_{\mu}$ the polarization vector of
$\rho$ meson and $g^2_{\rho \pi \pi}/4 \pi $ = 2.91
\cite{lixueqian}.

With the experimental branching ratios \cite{pdg2006} and above
formulae, we can obtain all relevant coupling constants as
summarized in Table~\ref{table}. Because
$\Delta^{++*}(1620)(\frac{1}{2}^-)$ has its mass below $K^+
\Sigma^+$ threshold, we take $g^2_{\Delta^*(1620) K \Sigma}/4
\pi=g^2_{\Delta^*(1620) N \pi}/4\pi=0.06$ from SU(3) relation.

\begin{table}
\caption{Relevant $\Delta^*$ paramters. \label{table}}
\begin{center}
\begin{tabular}{|ccccc|}
\hline Resonances  &Width & Decay  & Branching
& $g^2/4 \pi$  \\
&(MeV) & Channel & ratios (\%)&  \\
\hline $\Delta^*(1620)$ & 150 & $\pi N$ & 25 &0.06 \\
& & $\rho N$   &14 &0.37  \\
& & $K \Sigma$ &-  &0.06    \\
$\Delta^*(1920)$ & 200 & $\pi N$ &12.5 &0.18(GeV$^{-2}$) \\
& & $K\Sigma$ &2.1 &0.12(GeV$^{-2}$) \\

\hline
\end{tabular}
\end{center}
\end{table}

\subsection{Propagators}

In our amplitude calculation of Feynman diagrams in
Fig.~\ref{diagram}, we also need propagators for the pion,
$\rho^+$ meson, and the intermediate $\Delta^{++*}$ resonances
with half-integer spin. For the pion and the $\rho^+$ meson, the
propagators are :

\begin{equation}
G_{\pi}(q_{\pi})=\frac{i}{q_{\pi}^2-m^2_{\pi}},
\end{equation}

\begin{equation}
G^{\mu \nu}_{\rho}(q_{\rho})=- i (\frac{g^{\mu
\nu}-q^{\mu}_{\rho}q^{\nu}_{\rho}/q^2_{\rho}}{q^2_{\rho}-m^2_{\rho}})
\end{equation}
with $q_{\pi}$ and $q_{\rho}$ the four momenta of pion and
$\rho^+$ meson, respectively.  $m_{\pi}$ and $m_{\rho}$ are the
corresponding masses.

The propagators for the $\Delta^{++*}$ resonances with
half-integer spin can be constructed with their projection
operators and the corresponding Breit-Wigner
factor~\cite{liang02}. For the spin-(n+$\frac{1}{2}$), the
propagator can be written as
\begin{equation}
G^{n+\frac{1}{2}}_R(P_R)=P^{(n+\frac{1}{2})}\frac{2M_R}{P^2_R-M^2_R+iM_R\Gamma_R}
\end{equation}
where $1/(P^2_R-M^2_R+iM_R\Gamma_R)$ is the standard Breit-Wigner
factor; $M_R$, $P_R$ and $\Gamma_R$ are the mass, four momentum
and full width of the resonance, respectively.
$P^{(n+\frac{1}{2})}$ is the projection operator. Explicitly,
\begin{equation}
P^\frac{1}{2}(p)=\frac{\gamma \cdot p +M_R}{2M_R},
\end{equation}
\begin{equation}
P_{\mu\nu}^\frac{3}{2}(p)=\frac{(\gamma \cdot p + M_R
)}{2M_R}\left[ g_{\mu \nu} - \frac{1}{3} \gamma_\mu \gamma_\nu -
\frac{1}{3 M_R}( \gamma_\mu p_\nu - \gamma_\nu p_\mu) - \frac{2}{3
M^2_R} p_\mu p_\nu \right]
\end{equation}
for $\Delta^*(1620)(\frac{1}{2}^-)$ and
$\Delta^*(1920)(\frac{3}{2}^+)$, respectively.

\subsection{Amplitude and total cross section for $pp \to nK^+\Sigma^+$}

The full amplitude in our calculation for the $pp \to
nK^+\Sigma^+$ reaction is composed of three parts corresponding to
$\Delta^{++*}(1620)$ production from $\pi^+$ exchange,
$\Delta^{++*}(1620)$ production from $\rho^+$ exchange and
$\Delta^{++*}(1920)$ production from $\pi^+$ exchange,
respectively. Explicitly,
\begin{eqnarray}
{\cal M}={\cal M}(\Delta^{++*}(1620),\pi^+)+{\cal
M}(\Delta^{++*}(1620),\rho^+)+{\cal M}(\Delta^{++*}(1920),\pi^+).
\label{ampl}
\end{eqnarray}
Each amplitude can be obtained straightforwardly with effective
couplings and propagators given in previous sections by following
the Feynman rules. Here we give explicitly the amplitude $\cal
M$$(\Delta^{++*}(1620),\pi^+)$, as an example,
\begin{eqnarray}
{\cal M}(\Delta^{++*}(1620),\pi^+)& = &
\bar{u}_{\Sigma}(p_{\Sigma},s_\Sigma) g_{\Delta^* \Sigma K}
G_*(p_*) g_{\Delta^* N \pi} u_N(p_1,s_1)
G_{\pi}(q_{\pi})\times \nonumber\\
 && \bar{u}_N(p_n,s_n)\sqrt{2} g_{\pi NN} \gamma_5
u_N(p_2,s_2) \nonumber\\
 && + (\text {exchange term with } p_1 \leftrightarrow
p_2)
\end{eqnarray}
with $s_\Sigma$, $s_n$, $s_1$, $s_2$ the spin projection of
$\Sigma^+$, neutron in the final state and two initial protons,
respectively. $G_*(p_*)$ and $G_{\pi}$ are the propagators for
$\Delta^*(1620)$ resonance and the exchanges pion. $u_N$ and
$u_{\Sigma}$ are the Dirac wave functions of the nucleon and the
$\Sigma$. $p_{1}$ and $p_{2}$ represent the four momenta of the
initial protons.

Then the calculation of the invariant amplitude square $|\cal
M|^{\text {2}}$ and the cross section $\sigma (pp\to n K^+
\Sigma^+)$ is straightforward.
\begin{eqnarray}
d\sigma (pp\to n K^+ \Sigma^+)=\frac{1}{4}\frac{m^2_p}{F}
\sum_{s_1,s_2,s_n,s_\Sigma} |{\cal M}|^2\frac{m_n d^{3}
p_{n}}{E_{n}} \frac{d^{3} p_{K}}{2 E_{K}} \frac{m_{\Sigma} d^{3}
p_{\Sigma}}{E_{\Sigma}} \delta^4
(p_{1}+p_{2}-p_{n}-p_{K}-p_{\Sigma})  \label{eqcs}
\end{eqnarray}
with the flux factor
\begin{eqnarray}
F=(2 \pi)^5\sqrt{(p_1\cdot p_2)^2-m^4_p}. \label{eqff}
\end{eqnarray}
Since the relative phases between three parts in the amplitude of
Eq.(\ref{ampl}) are not known, the interference terms between
different parts are ignored in our concrete calculations.

\subsection{Final State Interaction}

For three particles in the final state, the interaction between
$K^+$ and $\Sigma^+$ is dominated by the s-channel $\Delta^{++*}$
resonances and is included in our resonance isobar model
calculation; the S-wave interaction between $K^+$ and nucleon is
known to be weak and repulsive \cite{Arndt} and does not play
significant role \cite{cosy06,05sibirt}. So we only need to
consider $n$-$\Sigma^+$ final state interaction.

In Ref.~\cite{06cosy11}, the total cross-section was measured at
the COSY-11 spectrometer at excess energies Q=13 MeV and Q=60 MeV.
At such near-threshold energies, the $n$-$\Sigma^+$ final state
interaction could play very significant role according to the
experience in studying a similar process $pp\to pK^+\Lambda$ both
experimentally \cite{cosy06} and theoretically \cite{05sibirt}. To
study possible influence from the $n$-$\Sigma^+$ final state
interaction, we include it in our calculation with the
Watson-Migdal approach~\cite{gold} by factorizing the reaction
amplitude as:
\begin{eqnarray}
A={\cal M}(pp \to n K^+ \Sigma^+) T_{n\Sigma} \label{fsiamp}
\end{eqnarray}
where ${\cal M}(pp \to n K^+ \Sigma^+)$ is the primary production
amplitude as discussed above,  $T_{n\Sigma}$ describes the
$n$-$\Sigma^+$ final state interaction, which goes to unity in the
limit of no FSI. For the near-threshold $n$-$\Sigma^+$ FSI, the
enhancement factor $T_{n\Sigma}$ can be expressed similarly as for
$p$-$\Lambda$ FSI in the study of $pp\to pK^+\Lambda$ process
\cite{05sibirt}:
\begin{eqnarray}
T_{n\Sigma}=\frac{q+ i \beta}{q- i \alpha}. \label{fsi}
\end{eqnarray}
where $q$ is the internal momentum of $n$-$\Sigma^+$ subsystem,
and the $\alpha$ and $\beta$ are related to the effective-range
parameters via
\begin{eqnarray}
a=\frac{\alpha + \beta}{\alpha \beta}, ~~~~~  r=\frac{2}{\alpha +
\beta}.
\end{eqnarray}

Lacking the knowledge on the $n$-$\Sigma^+$ interaction, in the
present work we assume similar $\alpha$ and $\beta$ values for
$n$-$\Sigma^+$ as for the $p$-$\Lambda$~\cite{05sibirt}, {\sl
i.e.},
\begin{eqnarray}
\alpha = - 75.0 ~~\text{MeV}, ~~~~~ \beta = 200.0 ~~\text{MeV}.
\end{eqnarray}

\section{Numerical results and discussion}

\begin{figure}[htbp]
\includegraphics[
height=3in, width=6.3in ]%
{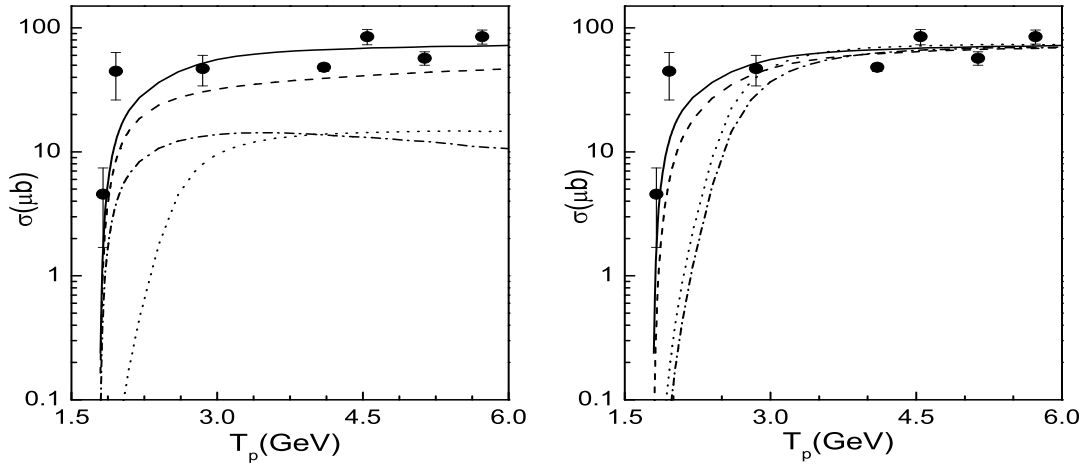}%
\caption{Total cross section vs T$_\text P$ for the $pp \to
nK^+\Sigma^+$ reaction from present calculation (solid curves)
compared with data~\cite{data,06cosy11}. Left: dot-dashed, dashed
and dotted curves for contributions from $\Delta^*(1620)(\pi^+$
exchange), $\Delta^*(1620)(\rho^+$ exchange) and
$\Delta^*(1920)(\pi^+$ exchange), respectively. Right:
contribution of effective $\Delta^*(1920)$ scaled by a factor 5
with FSI (dotted curve) and without FSI (dot-dashed); Full
calculation with FSI (solid) and without FSI
(dashed). } %
\label{tcs}%
\end{figure}

With the formalism and ingredients given above, the total cross
section versus the kinetic energy of the proton beam (T$_\text P$)
for the $pp \to nK^+\Sigma^+$ reaction is calculated by using a
Monte Carlo multi-particle phase space integration program. The
results for T$_\text P$ between 1.5 and 6.0 GeV are shown in
Fig.~\ref{tcs} together with experimental data
\cite{data,06cosy11} for comparison.

In the left figure of Fig.~\ref{tcs}, contributions from
$\Delta^*(1620)(\pi^+$ exchange), $\Delta^*(1620)(\rho^+$
exchange) and $\Delta^*(1920)(\pi^+$ exchange) are shown
separately by dot-dashed, dashed and dotted curves, respectively.
The contribution of the $\Delta^*(1620)$ production by the
$\rho^+$ exchange is found to be very important for the whole
energy range. The contribution of the $\Delta^*(1620)$ resonance
is found to be overwhelmingly dominant over $\Delta^*(1920)$
contribution for the two lowest data points close to the
threshold. This gives a natural source for the serious
underestimation of the near-threshold cross sections by previous
calculations \cite{tsushima,sibi,shyam06,gas}, which have
neglected either $\Delta^*(1620)$ resonance contribution
\cite{tsushima,sibi,shyam06} or $\rho^+$ exchange contribution
\cite{gas}. The solid curve in the figure is the incoherent sum of
the three contributions and reproduces the data quite well for the
whole energy range. The remained slightly underestimation for the
two lowest data points may be improved by the interference term
between $\rho^+$ exchange and $\pi^+$ exchange for the
$\Delta^*(1620)$ production.

To show the effect from the $n$-$\Sigma^+$ FSI, we give the result
without including the FSI factor by the dashed curve in the right
figure of Fig.~\ref{tcs}. Comparing dashed curve with the solid
curve which includes the FSI factor, we find that the FSI enhances
the total cross section by a factor of about 3 for the two lowest
data points. So the FSI is indeed making a significant effect at
energies close to threshold. But it does not change the basic
shape of the curve very much. In previous calculations
\cite{tsushima,sibi,shyam06}, only $\Delta^*(1920)$ contribution
are considered with a free scaling parameter to fit the data. In
Fig.~\ref{tcs} (right), we also show the results from only
$\Delta^*(1920)(\pi^+ $ exchange) scaled by a factor 5 for
comparison. It reproduces the data for T$_\text P$ above 2.8 GeV
quite well, but underestimates the two lowest data points by
orders of magnitude no matter whether including the FSI (dotted
curve) or not (dot-dashed curve).

\begin{figure}[htbp]
\includegraphics[
height=6in, width=6in]%
{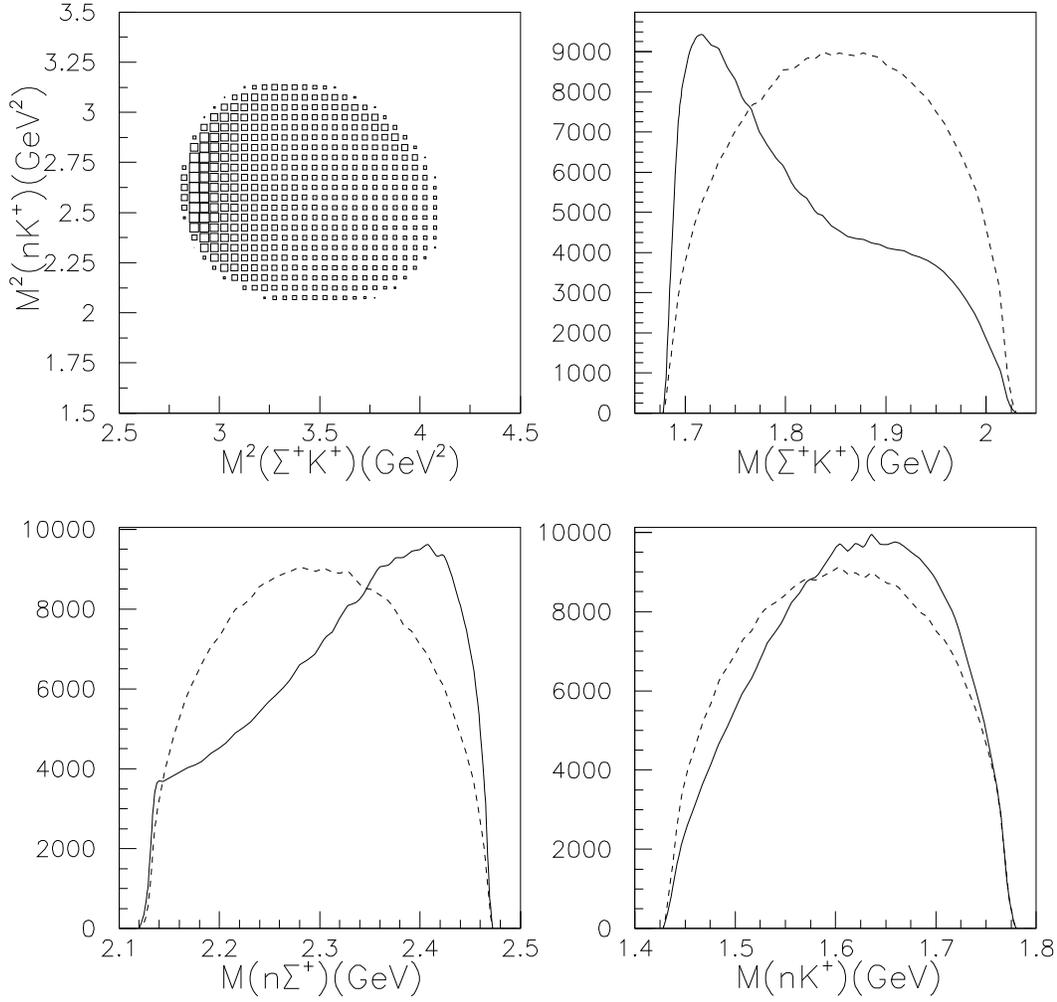}%
\caption{The Dalitz Plot and invariant mass spectra for the $pp
\to n K^+ \Sigma^+$ at T$_\text P$=2.8GeV, compared with pure
phase space distributions (dashed curves)}%
\label{imdp}%
\end{figure}

In Fig.~\ref{imdp}, we give our model prediction of the Dalitz
Plot and all relevant mass spectra for the reaction at T$_\text P$
= 2.8 GeV which can be achieved by the proton beams at Lanzhou
Cooler Storage Ring (CSR) and COSY at J\"ulich. A strong
$K^+\Sigma^+$ near-threshold enhancement is predicted together
with a relative weak broad peak around 1920 MeV in the
$K^+\Sigma^+$ invariant mass spectrum and some $n\Sigma^+$
near-threshold enhancement. The prediction may be checked by
future experiments at Lanzhou CSR with the scheduled $4\pi$ hadron
detector~\cite{menu2004} or at COSY with the newly installed
WASA-at-COSY detector~\cite{adam}.

The result on the dominant role of the $\Delta^{++*}(1620)1/2^-$
resonance for the near-threshold cross section of $pp \to
nK^+\Sigma^+$ reaction has many important implications.

(1) It gives another example that sub-threshold resonances can
make extremely important contributions and should not be simply
ignored. Many calculations were used to consider only the
resonances above threshold, such as previous calculations
\cite{tsushima,sibi,shyam06} for $pp \to pK^+\Lambda$ and $pp \to
nK^+\Sigma^+$. In the case of $pp \to pK^+\Lambda$, the
sub-$K^+\Lambda$-threshold resonance $N^*(1535)$ is proposed to
play a significant role for the near-threshold energies
\cite{liubc}.  Comparing with recent experimental measured Dalitz
plot by COSY-TOF Collaboration, a model fitting without including
the contribution of the $N^*(1535)$ visibly underestimates the
part near $K\Lambda$ threshold \cite{cosy}. The important role of
the $N^*(1535)$ for the $K\Lambda$ production is most clearly
demonstrated by the BES data on $J/\psi\to\bar pK^+\Lambda$
\cite{liubc,yanghx}. There are several more examples from $J/\psi$
decays showing the importance of contribution from sub-threshold
particles, such sub-$\pi N$-threshold nucleon pole contribution in
$J/\psi\to \bar pn\pi^+$ \cite{pnpi,liang2}, sub-$K\bar
K$-threshold contribution in $J/\psi\to K\bar K\pi$ and
sub-$\omega\pi$-threshold contribution in $J/\psi\to \omega\pi\pi$
\cite{wufq}.

(2) Since the $pp \to nK^+\Sigma^+$ and $pp \to pK^+\Lambda$
reactions are the basic inputs for the strangeness production from
heavy ion collisions \cite{Rafel,Li}, the inclusion of the
sub-threshold $\Delta^{*++}(1620)$ and $N^*(1535)$ contributions
may be essential for such studies.

(3) The t-channel $\rho$ exchange may play important role for many
meson production processes in proton-proton collisions and should
not be ignored.

(4) The extra-ordinary large coupling of the $\Delta^{*}(1620)$ to
$\rho N$ obtained from the $\pi^+ p\to N\pi\pi$
\cite{pdg2006,dytman} seems confirmed by our present study of the
strong near-threshold enhancement of $pp \to nK^+\Sigma^+$ cross
section. Does the $\Delta^{*}(1620)$ contain a large $\rho N$
molecular component or relate to some $\rho N$ dynamical generated
state~? If so, where to search for its SU(3) partners~? Sarkar et
al.~\cite{oset} have studied baryonic resonances from baryon
decuplet and psudoscalar meson octet interaction. It would be of
interests to study baryonic resonances from baryon octet and
vector meson octet interaction. In fact, from PDG compilation
\cite{pdg2006} of baryon resonances, there are already some
indications for a vector-meson-baryon SU(3) decuplet. While the
$\Delta^{*}(1620)1/2^-$ is about 85 MeV below the $N\rho$
threshold, there is a $\Sigma^*(1750)1/2^-$ about 70 MeV below the
$NK^*$ threshold and there is a $\Xi^*(1950)?^?$ about 60 MeV
below the $\Lambda K^*$ threshold. If these resonances are indeed
the members of the $1/2^-$ SU(3) decuplet vector-meson-baryon
S-wave states, we would expect also a $\Omega^* 1/2^-$ resonance
around 2160 MeV. All these baryon resonances can be searched for
in high statistic data on relevant channels from vector charmonium
decays by upcoming BES3 experiments in near future.

\bigskip
\noindent

{\bf Acknowledgements} We would like to thank H.C.Chiang,
P.N.Shen, B.C.Liu, H.Q.Zhou, F.Q.Wu and F.K.Guo for useful
discussions. This work is partly supported by the National Natural
Science Foundation of China under grants Nos. 10225525, 10435080,
10521003 and by the Chinese Academy of Sciences under project No.
KJCX3-SYW-N2.

\end{document}